\documentclass[usenatbib]{basi}
\usepackage[T1]{fontenc}
\usepackage[british]{babel}
\usepackage[varg]{txfonts}
\usepackage{natbib}
\conference{The Metrewavelength Sky}

\usepackage{dcolumn}
\begin{document}
\title[First results from CLoGS]{First Results from the Complete Local-Volume Groups Sample}
\author[E. O'Sullivan et~al.]%
       {E. O'Sullivan$^1$\thanks{email: \texttt{eosullivan@cfa.harvard.edu}},
       K. Kolokythas$^{2}$, S.~Raychaudhury$^{3,2}$, J. Vrtilek$^{1}$ and N. Kantharia$^{4}$\\
       $^1$Harvard-Smithsonian Center for Astrophysics, 60 Garden Street, Cambridge, MA 02138, USA\\
       $^2$School of Physics and Astronomy, University of Birmingham, Birmingham, B15 2TT, UK\\
       $^3$Department of Physics, Presidency University, 86/1 College Street, 700 073 Kolkata, India\\
       $^4$National Center for Radio Astrophysics, Tata Institute of Fundamental Research, Post Bag 3,\\ Ganeshkhind, 411 007 Pune, India}

\pubyear{2013}
\volume{00}
\pagerange{\pageref{firstpage}--\pageref{lastpage}}

\date{Received --- ; accepted ---}

\maketitle
\label{firstpage}

\begin{abstract}
Galaxy groups form the environment of the majority of galaxies in the local Universe, and many host an extended hot intra-group medium whose radiative cooling appears to fuel, and be stabilised by, feedback from AGN in group-central galaxies. Unfortunately studies of the physical properties of groups and the influence of AGN on their member galaxies and gaseous haloes have been limited by a lack of reliable representative samples of groups in the local Universe. To address this problem, we have assembled the Complete Local-Volume Groups Sample (CLoGS), an optically-selected statistically-complete sample of 53 groups within 80 Mpc, which we aim to observe in both low-frequency radio  and X-ray wavebands. We here describe results from the first half of the sample, for which X-ray and radio observations are complete. Roughly 55\% of the groups have group-scale X-ray halos, of which $\sim$65\% have cool cores a similar fraction to that found in galaxy clusters. While 25 of the 26 group central galaxies host radio AGN, among the X-ray bright groups only the cool core systems are found to support central jet sources.  
\end{abstract}


\section{Introduction}\label{s:intro}
Galaxy groups form the environment of the majority of galaxies in the local Universe. Many groups also host an extended hot intra-group medium (IGM) whose radiative cooling appears to fuel, and be stabilised by, AGN feedback from group-central galaxies. The group environment is conducive to galaxy evolution; low relative velocities and small galaxy separations drive mergers and tidal interactions, while motion through the IGM can produce gas compression or stripping. As such, groups are a critical environment in which to study the interplay between galaxy evolution, the development of the hot IGM, and AGN feedback. However, until recently, studies of groups have been hampered by the lack of reliable, statistically representative samples. Optically selected samples tend to become unreliable for low mass systems owing to the small number of detected group member galaxies. X-ray selection in the nearby Universe is dependent on the ROSAT All-Sky Survey, which is known to be biased toward centrally concentrated, cool core systems at low luminosities \citep{Eckertetal11}. This bias can be illustrated by comparing the cool-core fractions of clusters measured from statistical samples \citep[e.g., $\sim$50\%, ][]{Sandersonetal06} with the best available estimates from non-statistical samples of groups \citep[$\sim$85\%,][]{Dongetal10}. Without a reliable, statistical sample of nearby groups we cannot know whether this difference indicates a change in the IGM heating balance between groups and clusters, or whether it is merely a product of bias.

To resolve this problem, we have created a Complete Local-Volume Groups Sample consisting of 53 groups within 80~Mpc. We initially draw groups from the Lyon Galaxy Group sample \citep{Garcia93}, selecting systems with at least 4 members and at least one luminous early-type galaxy (L$_B>$3$\times$10$^{10}$L$_\odot$). Groups must also have Declination $>$-30$^\circ$, to ensure visibility from GMRT and VLA. We then refine and expand group membership, and exclude the richest and poorest groups, which either correspond to known clusters, or lack the numbers of galaxies needed for accurate estimation of properties such as velocity dispersion and spiral fraction. Further details of the selection can be found at \texttt{http://www.sr.bham.ac.uk/$\sim$ejos/CLoGS.html}.

Our goal is then to observe these optically selected groups in the X-ray, since the presence of a hot X-ray emitting IGM would confirm them as gravitationally bound systems, and in the radio, to allow examination of the AGN and star formation in the member galaxies. At present, we have observed the full 53-group sample with GMRT, using dual-frequency 235/610~MHz observations with a typical on-source time of 3-4 hours. In the X-ray we have observed a high-richness subsample of 26 groups, itself statistically complete, with a limiting sensitivity of L$_{0.5-7~keV}>$1.2$\times$10$^{42}$~erg~s$^{-1}$.

\section{Radio and X-ray detection rates}
In the 26-group high richness subsample, we detect 14 systems with diffuse X-ray emission extending $>$65~kpc and temperatures typical of galaxy groups (0.5-1.5~keV). A further 7 systems have smaller scale gaseous haloes associated with the dominant early-type galaxy. We therefore consider $\sim$55\% of this subsample to be confirmed as gravitationally bound groups. The sample is dynamically active, with two group-group mergers and two recently disturbed ``sloshing'' systems \citep[NGC~5044 and NGC~5846,][]{Gastaldelloetal13}. 

\begin{figure}
\centerline{
\includegraphics[width=7cm]{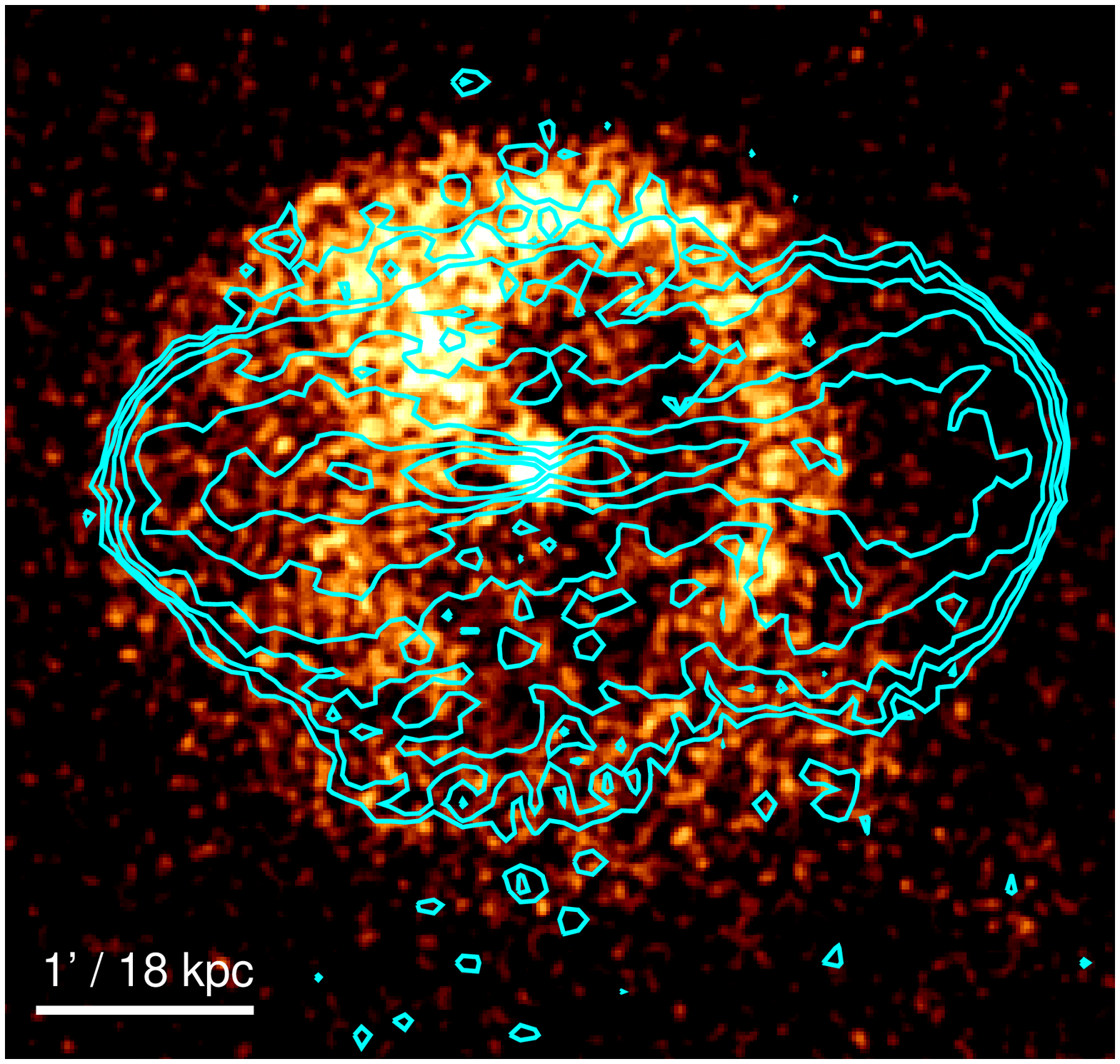}
\includegraphics[width=7cm]{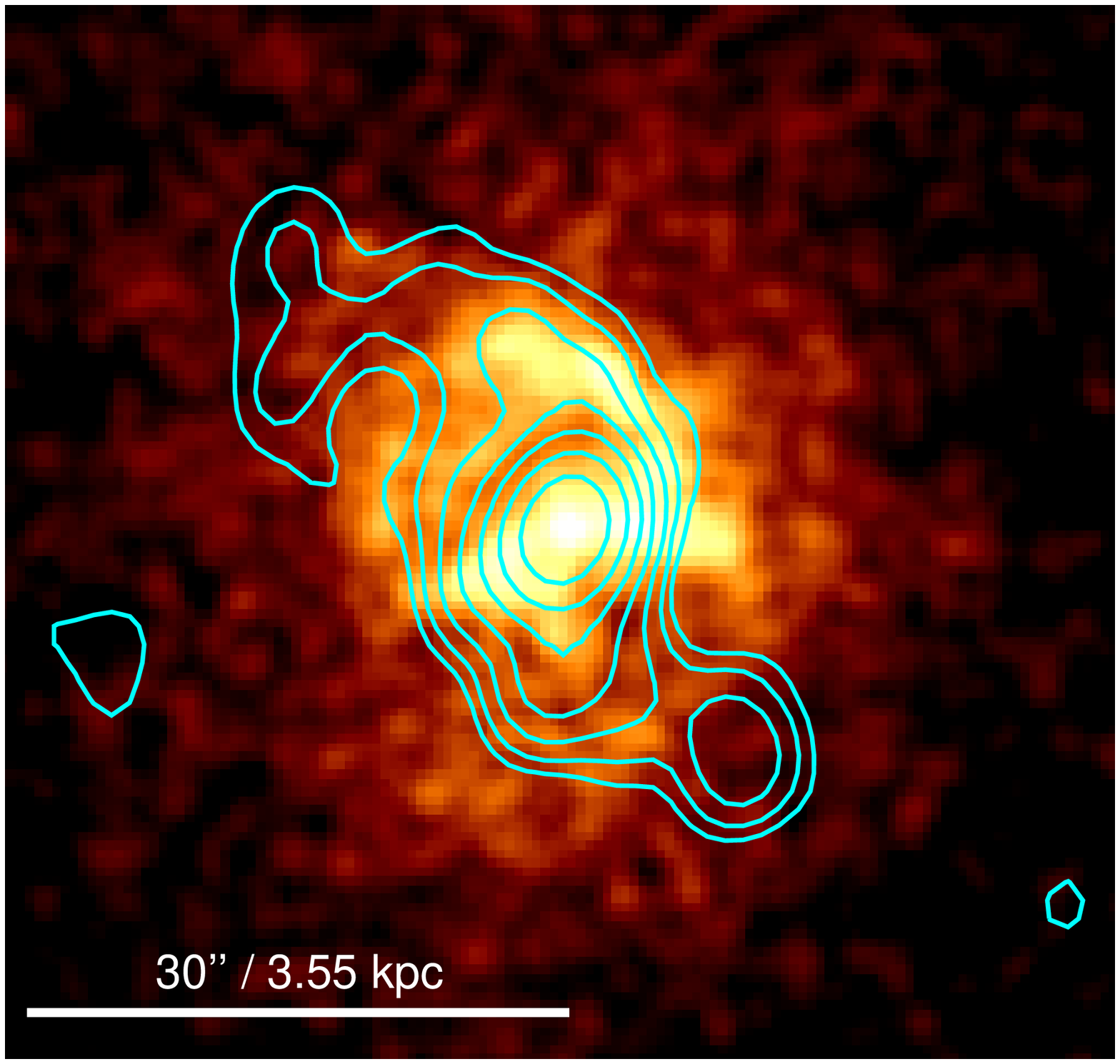}
}
\caption{\textit{Chandra} X-ray images of NGC 193 (\textit{left}) and NGC~5846 (\textit{right}) with radio contours overlaid. GMRT 610~MHz contours show the jets and cocoon of the FR-I radio galaxy 4C+03.01 in NGC~193, while VLA 1.4~GHz contours reveal a much smaller jet source in NGC~5846. In both cases the jets have opened cavities in the IGM, see \citet{Giacintuccietal11} for more details.\label{f.ims}}
\vspace{-0.1cm}
\end{figure}

We detect $\sim$45\% of the group member galaxies in the radio, with a higher detection rate ($\sim$70\%) for spiral galaxies than other morphologies ($\sim$25\%). This excludes the dominant early-type galaxies, whose location at the group centre makes them more likely to host AGN. We detect 25 of these dominant early-types at 235 and/or 610~MHz, and 7 host jet sources. The jet sources have a variety of morphologies, from small-scale jets $<$10~kpc long to large-scale FR-I radio galaxies such as 3C~270 and 4C+03.01 (see Fig.~\ref{f.ims}). In most cases we observe IGM cavities associated with the jets, and we estimate that the age and total energy outputs of the jets to be in the range 1-100~Myr and 10$^{55}$-10$^{59}$~erg. Further details of the radio properties of some of our groups are presented in \citet{Kolokythasetal14}.

\section{Cool core fraction and relation to the central AGN}
We define systems with a central decline in X-ray temperature profile of $>$3$\sigma$ significance as having cool cores. Previous studies have shown that cool cores are linked to the presence of H$\alpha$ emission and nuclear activity in cluster central galaxies \citep[e.g.,][]{Cavagnoloetal08} indicating that gas can cool from the X-ray phase to fuel the central AGN. We find that 9 of the 14 confirmed groups in our high richness subsample ($\sim$65\%) have cool cores, a similar fraction to that measured in clusters. Central galaxies with AGN jets are only found in cool core groups within our sample, in line with previous findings from more massive systems \citep{Sun09}.   

\begin{figure}
\centerline{
\includegraphics[width=7cm,bb=20 210 560 740]{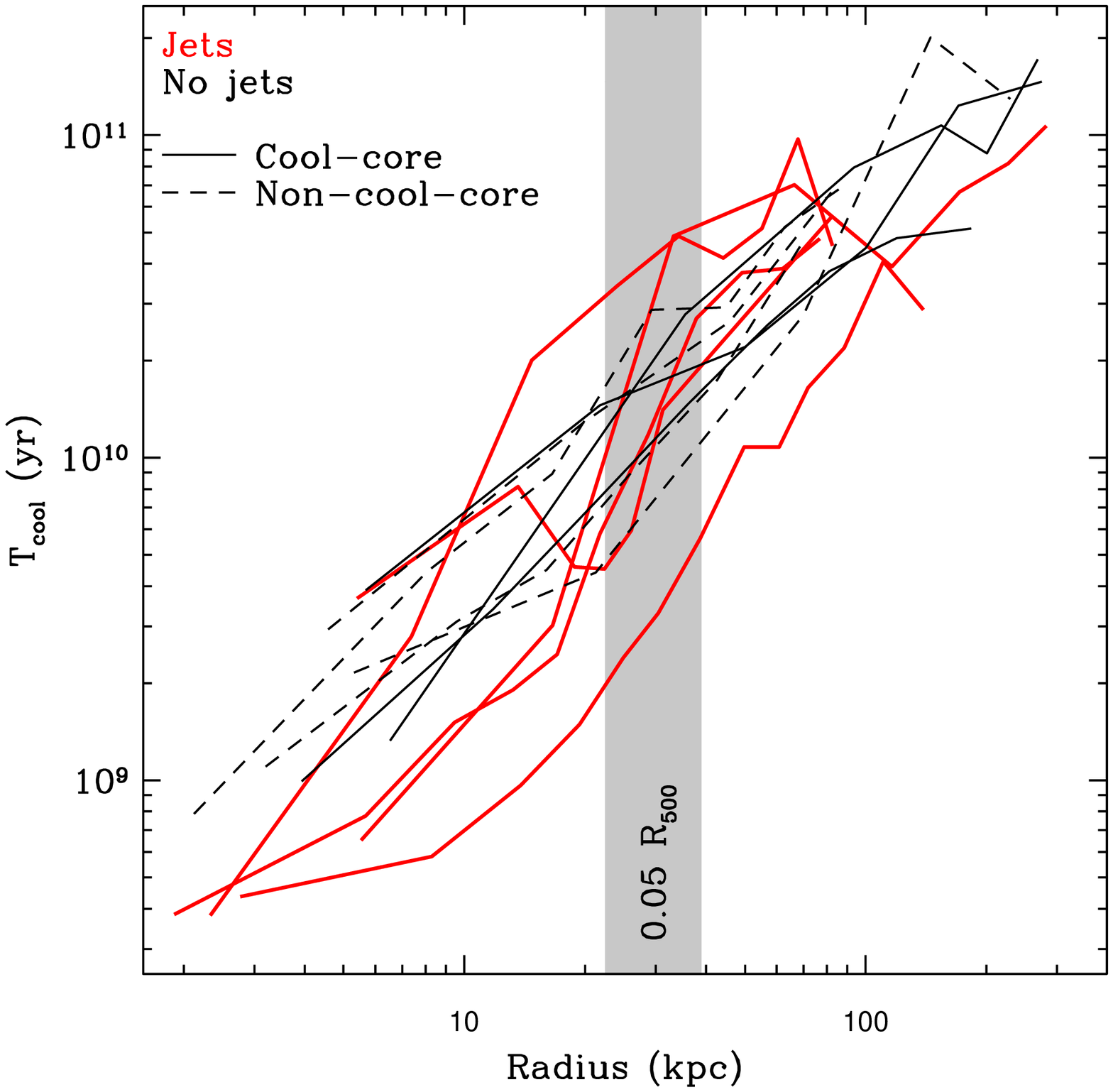}
\includegraphics[width=7cm,bb=20 210 560 740]{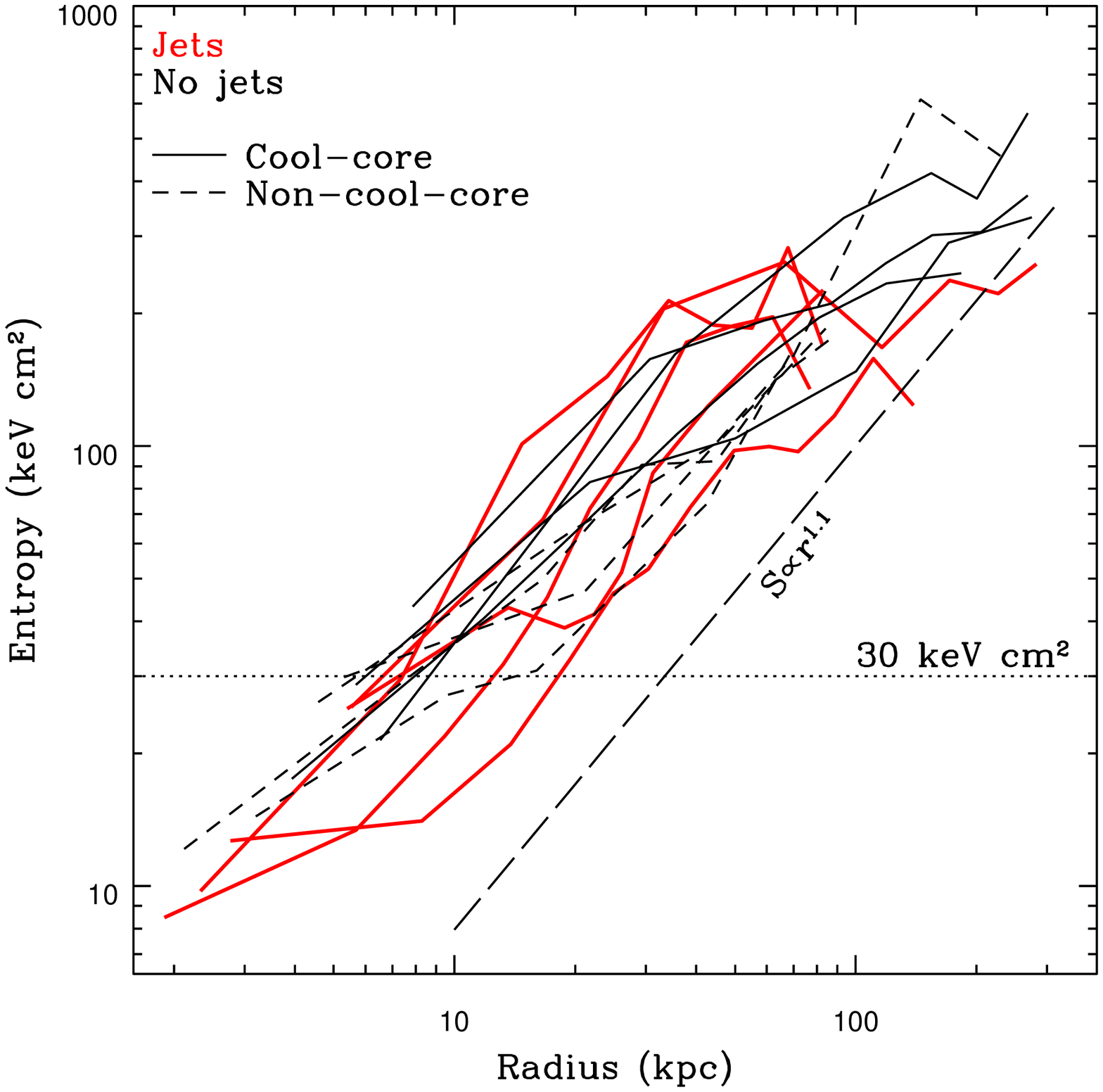}
}
\caption{Profiles of IGM cooling time (\textit{left}) and entropy (\textit{right}) for the X-ray bright CLoGS groups. The grey bar indicates the range of radii corresponding to 0.05 R$_{500}$ for these 0.5-1.5~keV systems. \label{f.profs}}
\end{figure}

Figure~\ref{f.profs} shows profiles of cooling time and entropy for our confirmed groups. The entropy profiles follow the canonical r$^{1.1}$ gradient in their outer parts, and many of the groups have central entropies below the 30~keV~cm$^2$ limit found to be associated with nuclear activity and H$\alpha$ nebulae. The groups also typically have short central cooling times of order 0.5-3~Gyr. \citet{Hudsonetal10} suggest a cool core classification scheme for clusters based on cooling time at 0.05 R$_{500}$, with strong cool cores (SCCs) having t$_{cool}<$1~Gyr and weak cool cores (WCCs) having t$_{cool}<$7.7~Gyr. Applying this to our groups, we would find no SCCs and only four WCCs, two of which have rising central temperature profiles. We therefore conclude that current nuclear jet activity is more closely linked to a declining central temperature profile than to the central entropy or cooling time in our groups.



\begin{thebibliography}{}
%
\bibitem[{{Cavagnolo} {et~al.}(2008){Cavagnolo} et al.}]{Cavagnoloetal08}
{Cavagnolo}, K.~W., {Donahue}, M., {Voit}, G.~M., \& {Sun}, M. 2008, ApJ, 683,
  L107

\bibitem[{{Dong} {et~al.}(2010){Dong}, {Rasmussen}, \& {Mulchaey}}]{Dongetal10}
{Dong}, R., {Rasmussen}, J., \& {Mulchaey}, J.~S. 2010, ApJ, 712, 883

\bibitem[{{Eckert} {et~al.}(2011){Eckert} et al.}]{Eckertetal11}
{Eckert}, D., {Molendi}, S., {Gastaldello}, F., \& {Rossetti}, M. 2011, A\&A, 526, A79

\bibitem[{Garcia(1993)}]{Garcia93}
Garcia, A.~M. 1993, A\&AS, 100, 47

\bibitem[{{Gasteldello} {et~al.}(2013){Gastaldello} et al.}]{Gastaldelloetal13}
{Gastaldello}, F., et al. 2013, ApJ, 770, 56

\bibitem[{{Giacintucci} {et~al.}(2011){Giacintucci} et al.}]{Giacintuccietal11}
{Giacintucci}, S., et al. 2011, ApJ, 732, 95

\bibitem[{{Hudson} {et~al.}(2010){Hudson} et al.}]{Hudsonetal10}
{Hudson}, D.~S., {Mittal}, R., {Reiprich}, T.~H., {Nulsen}, P.~E.~J.,
  {Andernach}, H., \& {Sarazin}, C.~L. 2010, A\&A, 513, A37

\bibitem[{{Kolokythas} {et~al.}(2014){Kolokythas} et al.}]{Kolokythasetal14}
{Kolokythas}, K., {O'Sullivan}, E., {Raychaudhury}, S., {Ishwara-Chandra}, C.~H. \& {Kantharia}, N.~G. 2014, proceedings of \textit{The Metrewavelength Sky}, arXiv:1402.5109.

\bibitem[{{Sanderson} {et~al.}(2006){Sanderson}, {Ponman}, \&
  {O'Sullivan}}]{Sandersonetal06}
{Sanderson}, A.~J.~R., {Ponman}, T.~J., \& {O'Sullivan}, E. 2006, MNRAS, 372,
  1496

\bibitem[{{Sun}(2009)}]{Sun09}
{Sun}, M. 2009, ApJ, 704, 1586
%
\end{thebibliography}

\label{lastpage}
\end{document}